\documentclass[journal]{IEEEtran}
\usepackage[pdftex]{graphicx}
\usepackage{amsmath}
\usepackage{amssymb}
\usepackage{blindtext}
\usepackage{color}

\ifCLASSINFOpdf
\else
\fi

\begin{document}

\title{Coupled Topological Surface Modes in Gyrotropic Structures: Green's Function Analysis}


\author{\IEEEauthorblockN{S. Ali Hassani Gangaraj, and Francesco Monticone~\IEEEmembership{Member,~IEEE}}

\thanks{Manuscript received xxx; revised xxx. 

The authors are with the School of Electrical and Computer Engineering, Cornell University, Ithaca, NY 14853 USA (e-mail: ali.ganagarj@gmail.com, francesco.monticone@cornell.edu).

}}

\markboth{Journal of \LaTeX\ Class Files,~Vol.~xxx, No.~xxx, xxx~2015}%
{Shell \MakeLowercase{\textit{et al.}}: Bare Demo of IEEEtran.cls for Journals}
%




\maketitle

\begin{abstract}

At a transition in a wave-guiding structure, part of the incident energy is transmitted and part of the energy is reflected. When the waveguide has non-trivial topological properties, however, the transition may occur with no backscattering, and with unusual modal coupling/transformations. Within this context, we discuss the response of a nonreciprocal topological structure composed of two nearby interfaces between oppositely-biased gyrotropic media and an isotropic medium, which support unidirectional surface modes (topological modes). We provide an exact Green's function analysis of this structure, and we discuss how the topological surface modes are modified when the two interfaces are brought closer and eventually merged. We show that the resulting mode conversion is independent of the transition geometry.

\end{abstract}


\maketitle

\IEEEdisplaynontitleabstractindextext

%
\IEEEpeerreviewmaketitle

\section{Introduction}

Topological materials and metamaterials are currently a topic of large interest in different domains of wave physics \cite{book_1}-\cite{Khanikaev}. Of particular relevance is their ability to support backscattering-immune surface waves on an interface with a material having different topological properties. The simplest realization of a topological material for electromagnetic waves is a nonreciprocal gyrotropic medium: when an interface with a conventional opaque medium is introduced, under certain conditions, a unidirectional surface mode emerges, completely immune to the presence of defects along its propagation \cite{Engheta_1}. While the study of unidirectional surface waves on gyrotropic materials dates back to the 1960s \cite{1962}, only recently it has been shown that the existence of these unidirectional surface modes originates from the non-trivial topological properties of the bulk eigenmodes of the gyrotropic medium \cite{Mario_1}-\cite{Hassani_ferrite}. 

Topology is a branch of mathematics that studies properties that are invariant under continuous deformations of the considered system. An example of a topological-invariant property is the \emph{genus} of a surface, which intuitively indicates the number of ``holes'' in the surface (a sphere has genus 0; a torus genus 1). Two surfaces having the same genus can be continuously deformed into one another. A different topological invariant that emerges in condensed-matter or electromagnetic systems is the \emph{Chern number}, which can often be interpreted as a ``winding number'' for the eigenmodal evolution in wavenumber space \cite{book_1}-\cite{Khanikaev}. 
%
Within this context, a key factor to realize topologically-non-trivial electromagnetic systems is to have a material with a band degeneracy in the bulk-mode dispersion diagram that is then lifted when a relevant symmetry of the system is broken (space-inversion, or time-reversal symmetry). In particular, if time-reversal symmetry is broken, a non-trivial bandgap opens, associated with a non-zero gap Chern number, which is the sum of the Chern numbers of all the modes below the bandgap, $ \mathcal{C}_{\mathrm{gap}} = \sum_{n<n_g} \mathcal{C}_n $ \cite{book_1}-\cite{Khanikaev}. This happens, for example, via the gyrotropy effect in biased plasmas or ferrites \cite{Mario_1}-\cite{Hassani_ferrite}, making these materials the electromagnetic equivalent of quantum-Hall electronic insulators \cite{book_1}. Opening or closing a non-trivial bandgap corresponds to a change of the global topological properties of a material. As a result, when two media with different topological properties share a common bandgap, at their interface the bandgap must completely close to allow for a modification of the gap Chern number from one medium to the other; therefore, surface modes emerge on the interface \cite{book_1}-\cite{Khanikaev_4}. In particular, the number of \emph{unidirectional} edge modes is equal to the difference of gap Chern numbers between the interfaced materials, a principle known as ``bulk-edge correspondence'' \cite{book_1}-\cite{Khanikaev}. 

\begin{figure}[bh!]
	\begin{center}
		\noindent \includegraphics[width=\columnwidth]{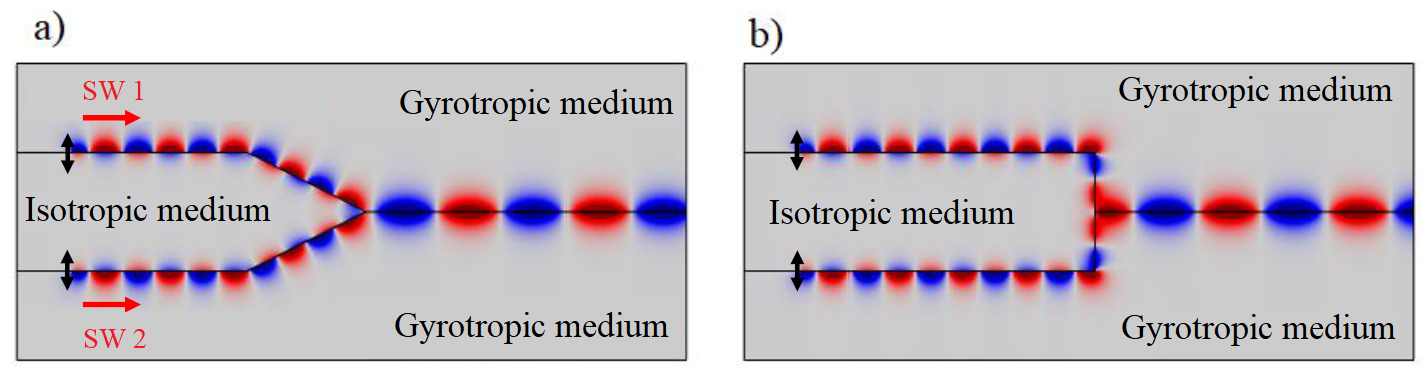}
	\end{center}
	\caption{Two examples of transition in a topological wave-guiding structure, leading to unusual coupling/transformation between unidirectional topological surface waves (SW). The plots show a time-snapshot of the vertical electric field distribution. The considered geometry is composed of two separate interfaces between an opaque isotropic medium with $ \epsilon_r = -1 $ (central layer) and two oppositely-biased gyrotropic media (plasmas) with plasma frequencies $ \omega_{p,1} = \omega_{p,2}  = 0.9 \omega  $ and cyclotron frequencies $  \omega_{c,1} = - \omega_{c,2}  = 0.28 \omega   $ where $ \omega $ is the source angular frequency. (a) Gradual transition; (b) abrupt transition. The bidirectional black arrows denote electric-dipole sources.}
	\label{2panel}
\end{figure}


Considering the case of gyrotropic materials mentioned above as a model for continuous electromagnetic topological insulators, it was reported in \cite{Hassani_1} that, at an interface between a magnetized plasma and a topologically-trivial opaque medium (i.e., $ \mathcal{C}_{\mathrm{gap}} =0 $), the difference of gap Chern numbers is unity, which predicts the existence of a single one-way surface mode. In this context, we have recently studied more complex situations, in which, for example, the topological surface mode is a leaky wave, hence obtaining nonreciprocal radiation effects \cite{leaky} (see also \cite{Khanikaev_1}), as well as the case of topological materials with gain and loss, which enable the emergence of peculiar branch-point-like modal degeneracies, known as ``exceptional points'' \cite{PRL_submitted}.

Another particularly interesting situation occurs when two topologically-protected surface modes, existing on different lossless interfaces, are forced to couple due to a change in wave-guiding geometry. This case is shown in Fig. \ref{2panel}: two oppositely-biased gyrotropic materials ($\mathcal{C}_{\mathrm{gap1}} = 1$ and $ \mathcal{C}_{\mathrm{gap2}} = -1 $) are separated by a topologically-trivial central layer ($\mathcal{C}_{\mathrm{gap}} = 0$), whose thickness is reduced gradually (Fig. \ref{2panel}a) or abruptly (Fig. \ref{2panel}b), such that the two material interfaces are brought closer to each other until they merge into a single interface. On this common interface, the jump in gap Chern number becomes $ \mathcal{C}_{\mathrm{gap1}}  - \mathcal{C}_{\mathrm{gap2}} = 2 $, which predicts the existence of two topological edge modes. Interestingly, our numerical simulations in Fig. 1 show that, when two unidirectional surface waves are separately launched on the two original interfaces, they seamlessly merge into a single surface wave with different wavenumber. Due to its topological nature, this modal transformation occurs with zero reflection even in the case of abrupt step transition in Fig. 1(b). Moreover, it seems that, in both cases considered in Fig. 1, only one of the two available modes of the common interface is excited by the waveguide transition. This behavior raises interesting questions: why is the second surface mode predicted by the bulk-edge correspondence not excited at all? How does the waveguide transition control this topological modal transformation? In order to answer these questions rigorously (not based on perturbation methods such as coupled-mode theory), in this Letter we develop an exact electromagnetic Green's function analysis of the considered wave-guiding structure.

\begin{figure}[bh!]
	\begin{center}
		\noindent \includegraphics[width=0.9\columnwidth]{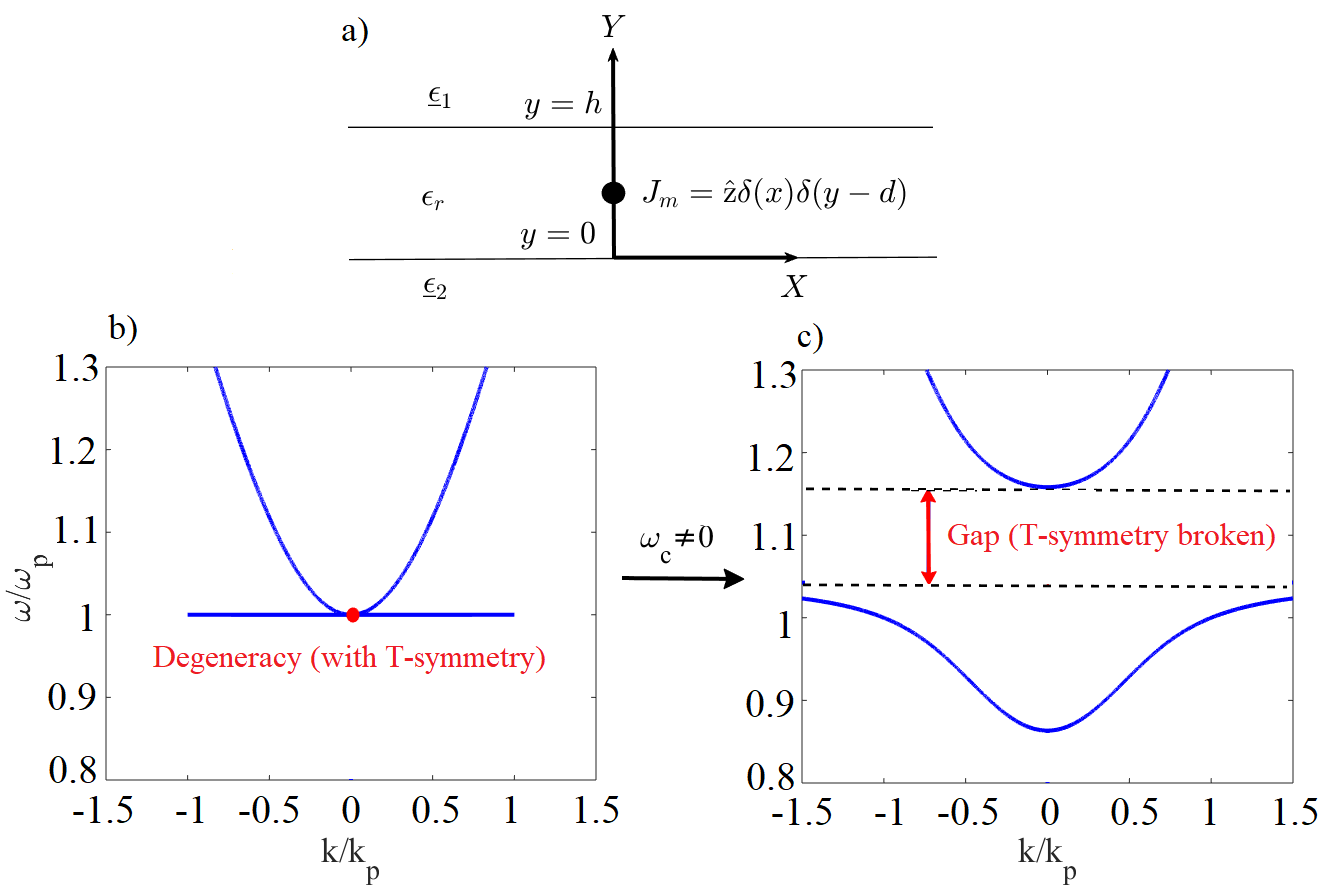}
	\end{center}
	\caption{(a) Geometry considered for the Green's function calculation: a two-dimensional multi-layered structure with a magnetic line source embedded in the central isotropic region. (b-c) Transverse-magnetic bulk-mode dispersion diagram for a plasma with (b) $  \omega_{c}  = 0 $ (unbiased) and (c) $ \omega_c / \omega_p = 0.28 $ (biased in the $z$ direction; wave propagation orthogonal to the bias). }
	\label{TM_bulk}
\end{figure}
 
 \section{Green's function and dispersion equation}
 
As shown in Fig. \ref{TM_bulk}(a), we consider a lossless multi-layered structure with interfaces normal to the $y$-axis and translationally invariant in the $x-z$ plane. An isotropic homogeneous material with permittivity $ \epsilon_r $ occupies the region $ 0<y<h $, and two different gyrotropic media described by permittivity and permeability tensors $ \underline{\boldsymbol{\epsilon}}_i = \epsilon_0 \left( \epsilon_{t,i} \boldsymbol{\mathrm{I}}_t + \epsilon_{a,i} \hat{\boldsymbol{\mathrm{z}}} \hat{\boldsymbol{\mathrm{z}}} + i \epsilon_{g,i} \hat{\boldsymbol{\mathrm{z}}} \times \boldsymbol{\mathrm{I}}   \right) $, $ \underline{\boldsymbol{\mu}}_i = \mu_0 \boldsymbol{\mathrm{I}} $, $ i=1,2 $, occupy the regions $ y > h $ and $ y<0 $ respectively, where ${\boldsymbol{\mathrm{I}}}_{t}=\boldsymbol{\mathrm{I}}-\mathbf{{\hat{z}}{\hat{z}}}$ ($ \boldsymbol{\mathrm{I}}  $ is the unit tensor). 

 
A magnetic line source, $ \boldsymbol{\mathrm{J}}_m = \hat{\boldsymbol{\mathrm{z}}} \delta(x) \delta(y-d) $, is assumed to exist in the dielectric region. Considering this $z$-invariant geometry and excitation, for the transverse-magnetic mode of the structure (TM mode; magnetic field along $z$), the radiated magnetic field $ H_z (x,y) $ satisfies the following wave equation 
%
\begin{equation}\label{Hz_GF}
\left[  \partial^2 / \partial x^2 + \partial^2 / \partial y^2 + k_0^2 \epsilon_r   \right] H_z(x,y) = -i\omega \epsilon_0 \epsilon_r \delta(x) \delta(y-d)
\end{equation}
where $ k_0 = \omega/ c $ and $c$ is the vacuum speed of light. The $x$-invariance of the geometry suggests the following integral representation for the Green's function $ H_z(x,y) $
\begin{equation}\label{FT_Hz}
H_z(x,y) = \frac{1}{2 \pi} \int_{-\infty}^{ + \infty } \overline{H}_z (k_x, y ) e^{ik_x x} dk_x
\end{equation}
where $ \overline{H}_z (k_x, y ) $ is the spatial spectrum (Fourier transform) of the magnetic field in wavenumber space, and $k_x$ is the wavenumber (propagation constant) parallel to the interfaces. It follows from (\ref{Hz_GF}) and (\ref{FT_Hz}) that
\begin{equation}\label{FT_GF}
\left[  \partial ^2 / \partial y^2 + k_y^2  \right] \overline{H}_z (k_x, y ) = -i\omega \epsilon_0 \epsilon_r \delta(y-d)
\end{equation}
with $ k_y= \sqrt {k_0^2 \epsilon_r - k_x^2} $. The solutions of (\ref{FT_GF}) in the central region are of the form $ \overline{H}_z(k_x, y ) = B ~ \cos(k_y y) + C~ \sin(k_yy), ~ 0 < y < d $, and $ \overline{H}_z(k_x, y ) = E ~ \cos(k_y y) + F~ \sin(k_yy),~ d < y < h $, with the jump condition at the source point $  \partial_{y} \overline{H}_z(k_x, d^+ ) -  \partial_y \overline{H}_z(k_x, d^ - ) = -i \omega \epsilon_0 \epsilon_r $, where $B, ~C, ~ E $ and $F$ are unknown field amplitudes. 

In order to apply boundary conditions at the two material interfaces, one has to make some assumptions for the field distribution inside the gyrotropic layers. As mentioned earlier, topologically-protected surface modes appear when we operate in the bandgap of the bulk modes of the gyrotropic material, i.e., the bandgap obtained by breaking time-reversal symmetry in Fig. 2(c).  
%
In reciprocal media and structures exhibiting a frequency gap, such as in periodic structures near the Bragg condition, the bandgap is often trivial, which means that the bands below/above the gap cannot be associated to any topological invariant number. Conversely, when a non-trivial bandgap is opened (e.g., in the bulk of a biased plasma), the eigenmode profile exhibits a topologically-non-trivial ``rotation'' as a function of wavevector
, leading to a non-zero winding number in momentum space, i.e., non-zero Chern number \cite{Khanikaev}. 
Interestingly, among different configurations of bulk-mode propagation in a biased plasma, only the TM mode propagating orthogonal to the bias provides non-trivial topological properties \cite{Hassani_2}. The dispersion equation of this mode is $ k_x^2 + k_y^2 = \epsilon_{eff} ( \omega / c )^2 $, where $ \epsilon_{eff} = ( \epsilon_{t}^2 - \epsilon_{g}^2 ) / \epsilon_{t} $ is the effective permittivity. The solutions of this dispersion equation are shown in Fig. \ref{TM_bulk}(b),(c): when there is no bias, a degeneracy occurs  at $ k= \sqrt{k_x^2 + k_y^2} = 0 $ and $ \omega = \omega_p $; once we turn on the bias, the degeneracy is lifted and a non-trivial bandgap opens in the dispersion diagram. In this case, the gap Chern number becomes non-zero (the details are provided in \cite{Mario_1,Hassani_2}), and a unidirectional surface mode is expected to emerge at frequencies within the gap.
%

For a generic TM surface mode existing on one of the interfaces of the structure in Fig. 2, we consider vertically decaying fields inside the gyrotropic layers with attenuation constant $ \alpha_{pi} = \sqrt{ k_x^2 - k_0^2 \epsilon_{eff,i}  }, ~ i=1,2$, such that $ \mathrm{Re}(\alpha_{pi}) > 0 $. By applying the boundary conditions for the continuity of tangential fields at $ y=0 $ and $ y= h $, one can find that   
\begin{align}\label{Exact_Hz}
	& H_z(x,y) = \frac{-i\omega \epsilon_0 \epsilon_r}{2\pi} \int_{-\infty}^{+\infty} dk_x  \frac{\cos(k_yd) +\gamma_3 \sin(k_yd)   }{  k_y(\gamma_3 - \gamma_1) } \times \notag \\& ~~~~~~~~~~~~~~~ \left( \cos(k_y y) + \gamma_1 ~ \sin(k_y y)  \right) e^{ik_xx} , ~~  0 < y < d \notag \\& 
	H_z(x,y) = \frac{i\omega \epsilon_0 \epsilon_r}{2\pi} \int_{-\infty}^{+\infty} dk_x  \frac{\cos(k_yd) +\gamma_1 \sin(k_yd)   }{  k_y(\gamma_3 - \gamma_1) } \times \notag \\& ~~~~~~~~~~~~~~ \left( \cos(k_y y) + \gamma_3 ~ \sin(k_y y)  \right) e^{ik_xx}, ~~  d < y < h
\end{align}
with the following parameters
\begin{align}\label{gamma}
	&\gamma_1 = \frac{\epsilon_r}{ k_y} \frac{\alpha_{p2} k_x - k_0^2 \epsilon_{g2}   }{ \epsilon_{t2} k_x + \epsilon_{g2} \alpha_{p2}  },~ \gamma_2 = \frac{\epsilon_r}{ k_y} \frac{-\alpha_{p1} k_x - k_0^2 \epsilon_{g1}   }{ \epsilon_{t1} k_x - \epsilon_{g1} \alpha_{p1}  }, \notag \\&
	\gamma_3 = \frac{ N_2 ~ \cos(k_y h) + D_2 \sin(k_yh) }{ D_2 \cos(k_yh ) - N_2 \gamma_2 ~ \sin(k_yh) }.
\end{align}

In the central region, the integrals in (\ref{Exact_Hz}) can be evaluated approximately as a sum of residue terms corresponding to the different surface modes of the structure, $H_z(x,y) = 2 \pi i \sum_m  w^{\mathrm{spp}} (k_{mx}, \omega) e^{ik_{mx}x}$, where $ w^{\mathrm{spp}}(k_x, \omega) = \mathcal{N}(k_x, \omega)/\partial_{k_x} \mathcal{D}(k_x, \omega) $ and
\begin{align}
	& \mathcal{N}(k_x, \omega) = -i \omega \epsilon_0 \epsilon_r D_1 D_3 \left ( D_3 ~ \cos(k_yd) +N_3~\sin(k_yd)   \right) \times \notag \\& ~~~~~~~~~~~~~~ \left( D_1 ~ \cos(k_yd) +N_1~\sin(k_yd)   \right) \notag \\&
	\mathcal{D}(k_x, \omega ) =  \left[D_1D_3 k_y \left(N_3 D_1 - N_1 D_3    \right)\right]
\end{align}
where $ N_i $ and $ D_i $, $i=1,3$, are the numerators and denominators of $ \gamma_i$. $ k_{mx} $ is the $m$th root ($m$th Green's function pole) of $ \mathcal{D}(k_x, \omega) = 0 $, which is the \emph{dispersion equation} of the unidirectional surface modes, for a given set of parameters. 

This formulation allows investigating how the topological modes evolve as we bring the two interfaces in Fig. \ref{TM_bulk}(a) closer. 

 
\section{Coupled topological modes}

\subsection{Coupling through an opaque medium}



\begin{figure}[h!]
	\begin{center}
		\noindent \includegraphics[width=0.9\columnwidth]{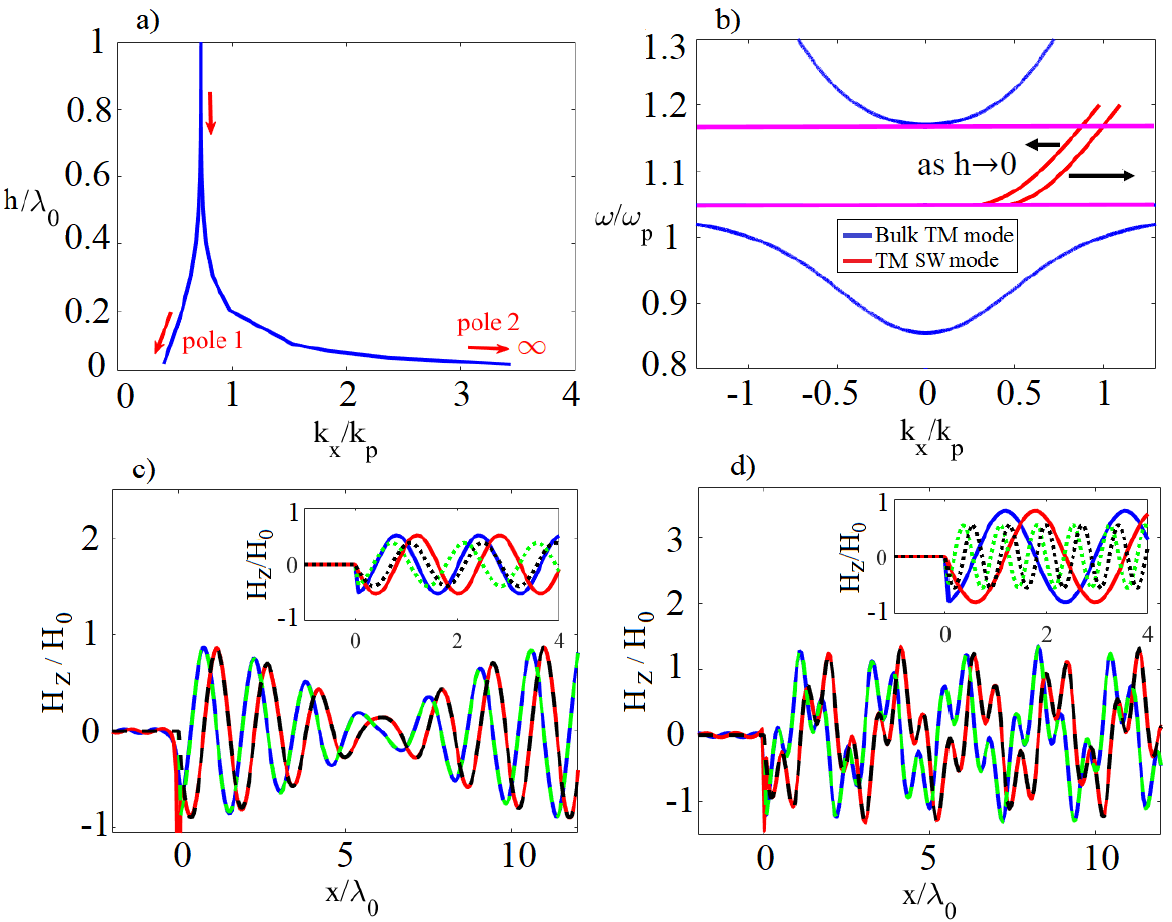}
	\end{center}
	\caption{(a) Evolution of the two unidirectional TM surface modes (poles of Green's function), for the geometry in Fig. 2, as a function of thickness of an opaque ($ \epsilon_r = -1 $) isotropic spacer. The gyrotropic media have the same properties as in Fig. 1. $ \lambda_0 $ is the free-space wavelength at the source frequency. (b) Dispersion bands for the bulk TM modes of the biased gyrotropic media (solid blue) and for the unidirectional TM surface modes (solid red), for a spacer thickness of $ h = 0.63 \pi c /\omega_{p}$. The two horizontal purple lines indicate the lower/upper edges of the bulk-mode bandgap. Black arrows indicate the movement of the bands as $h$ is reduced. (c) Magnetic field distribution in an opaque isotropic spacer of thickness $ h = 0.4 \lambda_0 $, for a source at $x=0, y = d = 0.7 h$. Solid blue and red lines are the real and imaginary parts of $ H_z(x,y= 0.9d) $ obtained from the exact solution, Eq. (\ref{Exact_Hz}). Dashed green and black lines are the real and imaginary parts obtained from residue evaluation. Inset shows the field distribution for the two modes separately. Solid and dotted lines indicate the real and imaginary parts of the first and second modes. (d) Same as (c) but for $ h = 0.1 \lambda_0 $.}
	\label{fig1}
\end{figure}

When the two interfaces in Fig. \ref{TM_bulk}(a) are sufficiently far from each other, and the spacer layer is opaque, the surface waves supported by the individual interfaces decay fast inside the central region, such that there is no coupling between them. If the two gyrotropic materials have the same properties ($ \omega_{p1} =\omega_{p2} =  \omega_p $) and equal but opposite biasing fields $ \omega_{c1} = - \omega_{c2} = \omega_c $, the structure supports two identical topologically-protected and unidirectional surface modes, localized on each interface separately without any interference. The coupling between the two modes begins by decreasing the thickness of the opaque spacer. Fig. \ref{fig1}(a) shows the evolution of the two solutions of the dispersion equation $ \mathcal{D}(k_x, \omega) = 0 $, at a given frequency, as the gap thickness decreases. For large gap thicknesses, the two uncoupled surface waves have the exact same propagation constant. For smaller thicknesses, the coupling perturbs the system, and the surface-wave solutions are ``forced apart'', forming two distinct unidirectional surface modes. In the limit of $ h \rightarrow 0 $, one of the solutions converges to a positive and finite value of $k_x$, while the second solution asymptotically tends to infinity. 

Fig. \ref{fig1}(b) shows the dispersion diagram of the structure in Fig. 2(a) with an opaque spacer of finite thickness. The solid blue lines indicate the TM bulk modes of the plasmas. Our analysis in Section II yields two additional dispersion curves crossing the entire bandgap, indicating the presence of two unidirectional topological surface modes with positive group velocities $ v_g = \partial \omega / \partial k_x >0  $. A clear consequence of nonreciprocity and topological protection is the fact that, in this non-trivial bandgap, there are no solutions of the dispersion equation for negative $k_x$, which results in the absence of any backward-propagating mode, and zero backscattering at discontinuities. If the opaque gap thickness is reduced, the dispersion line of the first surface mode moves toward left, converging to a finite values of $k_x>0$ for zero thickness, whereas the second dispersion line shifts rapidly toward positive and very large values, $k_x \to \infty$. 

These findings answer one of the questions raised in the Introduction: when the two original interfaces merge into a single interface between ``opposite'' gyrotropic media, there are still two unidirectional surface modes, in agreement with the bulk-edge correspondence; however, one of the two modes has extremely large (in theory, infinite) wavenumber $k_x$, which leads to a very large wave-impedance for this TM mode, $Z_{TM} \propto k_x$. For this reason, the mode is completely impedance-mismatched from the surface modes existing before the transition region in Fig. 1, hence the waveguide transition will not excite this mode at all. This fact, combined with the absence of backscattering due to topological protection, forces all the energy from the two incoming surface modes to be transfered to a single unidirectional mode, independently of the geometry of the waveguide transition.

Figure \ref{fig1}(c) shows the field distribution for a separation $ h = 0.4 \lambda_0 $, at which the two unidirectional surface waves are weakly coupled and have similar wavenumber. 
As clearly seen in Fig. \ref{fig1}(c), there is no propagation toward the negative $x$-axis, confirming the unidirectional nature of the topological modes. The two surface waves form an interference pattern as they propagates along the positive $x$-axis. The inset shows the two unidirectional surface waves separately, obtained from the residue calculation: the two modes have similar wavelengths and are excited with slightly different amplitudes. Fig. \ref{fig1}(d) shows the case of smaller gap thickness $ h = 0.1 \lambda_0 $, at which the difference in wavenumber of the two surface waves is larger, leading to a different interference pattern. As seen in the inset, the second mode oscillates faster and has smaller amplitude. By decreasing the spacer thickness further, the amplitude of the highly-oscillating mode becomes smaller and smaller; in the limit of $ h \rightarrow 0 $, the slowly-oscillating mode completely dominates the field distribution. The second mode still exists but with a vanishing wavelength and amplitude.

\subsection{Coupling through a transparent medium}

\begin{figure}[tb!]
	\begin{center}
		\noindent \includegraphics[width=0.9\columnwidth]{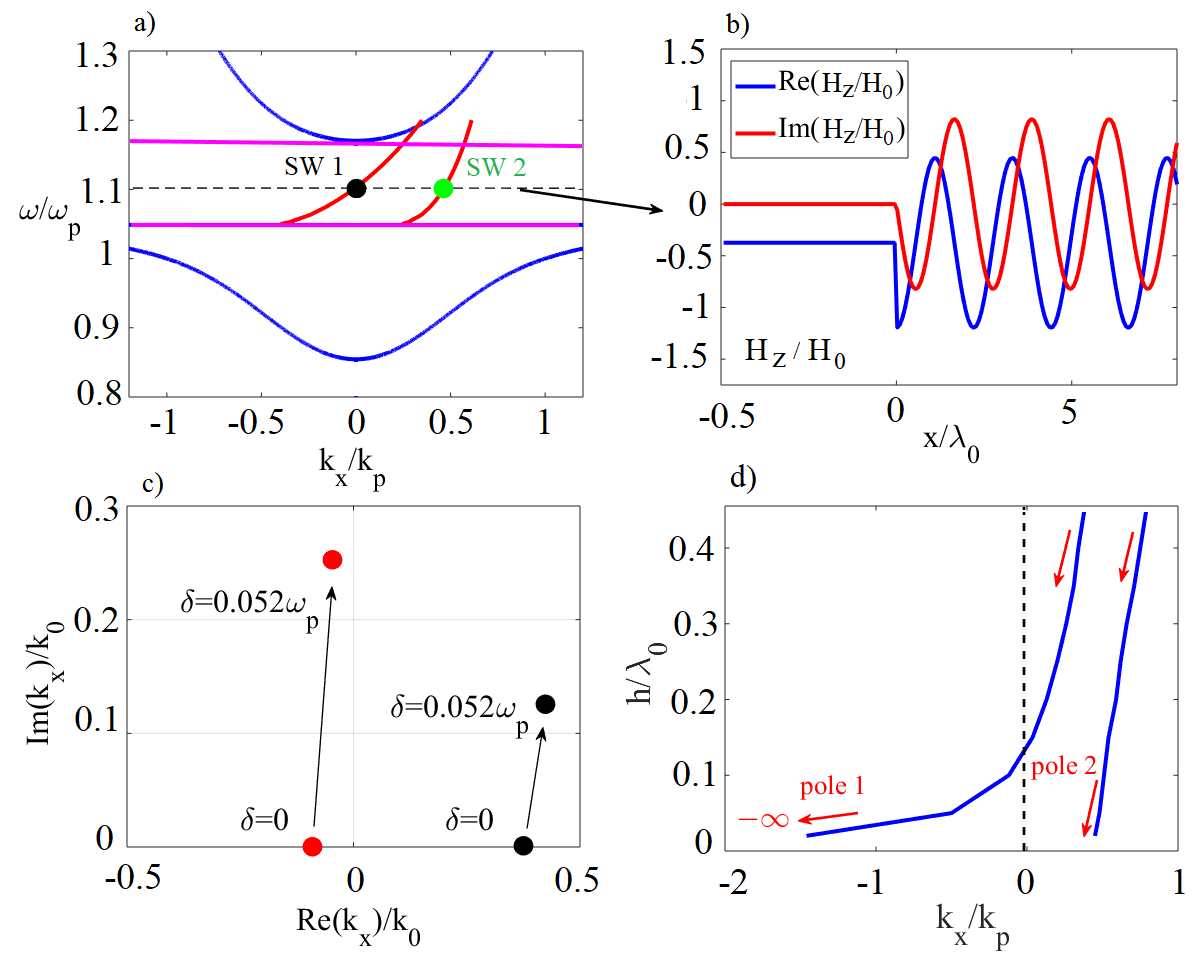}
	\end{center}
	\caption{(a) Dispersion diagram, similar to Fig. 3(b) but for a transparent dielectric spacer with $ \epsilon_r = 1 $ and $ h = 0.36 \pi c/ \omega_p  $. (b) Magnetic field distribution, $ H_z(x,y=0.9d) $, calculated by residue evaluation, for a source at $x=0, y = d = 0.7 h$. (c) Locus of the two surface-wave poles in the complex wavenumber plane, for $ \omega/\omega_p = 1.078  $, from lossless to lossy case. (d) Similar to Fig. 3(a), but for a transparent isotropic spacer with $ \epsilon_r = 1 $ (for this case, we assumed $ \omega / \omega_p = 1.13 $, $ \omega_c / \omega_p = 0.31 $).}
	\label{fig2}
\end{figure}

Topological mode coupling through a transparent dielectric gap with $ \epsilon_r >0 $ leads to a different modal behavior. The dispersion curves of the two topological surface modes of this structure are shown in Fig. \ref{fig2}(a), considering vacuum as the transparent spacer. Interestingly, in this situation the dispersion band of one of the surface modes crosses the vertical axis at $  k_x = 0 $ (black dot), while the other dispersion band is entirely on the positive side of the wavenumber axis. 

By denoting the frequency of this crossing point as $\omega_{cp}$, we can identify different cases: (i) For frequencies $\omega>\omega_{cp}$, the two surface waves propagates unidirectionally along the positive $x$-axis and interfere, similar to the opaque-spacer case. (ii) At $\omega=\omega_{cp}$, a unidirectional $k_x-$positive surface wave ``interferes'' with a surface wave with diverging phase velocity and wavelength, which leads to a spatially-constant shift in the real part of the field distribution excited by a line source, as shown in Fig. \ref{fig2}(b). (iii) For $\omega<\omega_{cp}$, there are two solutions with opposite sign of $ k_{x} $; the phase velocity of the first mode is negative, $ v_p = \omega/k_x < 0 $, whereas the slope of the dispersion curve is positive, hence the group velocity $ v_g = \partial \omega / \partial k_x >0 $. In this case, the mode is still unidirectional, and if the source is located at $x=0$, the excited fields are zero for $x<0$ (if the mode existed for $x<0$, the positive value of $v_g$ would indicate energy flowing toward the source, an unphysical situation). The correct interpretation of this case is that the excited surface wave propagates toward right with positive group velocity (energy flows away from the source) but with negative phase velocity. To validate this interpretation, Fig. \ref{fig2}(c) shows the dispersion equation solutions, at $ \omega/\omega_p = 1.078<\omega_{cp}$. When the plasmas are lossless, we have two solutions on the real axis of the wavenumber complex plane (red and black circles). 
By introducing a small level of loss into the plasmas (collision frequency $ \delta = 0.052 \omega_p $), the two solutions migrate toward the upper half-plane, 
which is indeed the proper Riemann sheet for right-going decaying surface modes. 

Finally, we show in Fig. \ref{fig2}(d) the evolution of the two solutions of the dispersion equation varying the transparent spacer thickness, similar to Fig. \ref{fig1}(a). Analogously to the case of opaque spacer, when $ h \rightarrow 0 $, one of the poles moves toward left, converging to a finite value of $k_x>0$, whereas the other pole moves to the negative $ k_x-$axis, and asymptotically tends toward negative infinity. Different from Fig. \ref{fig1}(a), however, if the thickness is increased, the two poles move closer to each other but do not converge. 
For larger values of $h$, the structure supports additional waveguide modes in the central layer.
	
\section{Conclusion}

We have presented an exact Green's function analysis of coupled topological surface modes in a wave-guiding structure composed of oppositely-biased gyrotropic media separated by an isotropic layer. 
Our analysis provides relevant insight into a broad class of topologically-protected modal transitions. Our theoretical predictions may be experimentally tested using different platforms. For example, a popular solution to realize gyrotropic materials at THz frequencies is to use $n$-type semiconductors, such as Indium Antimonide (InSb), under a magnetic bias. To give some numbers, the plasma frequency of a typical InSb sample is $ \omega_p / 2\pi \approx  5$  THz and collision frequency $ \delta / 2 \pi \approx 0.5 $ THz \cite{Palik}. Considering a feasible bias field up to $B_0=4~ T$, it is possible to open a moderately-wide bandgap in the range 6.8-8 THz, sufficient to observe the described effects in a suitably designed system. Our results and considerations, while derived for the case of biased gyrotropic materials, are expected to qualitatively apply to any nonreciprocal topological wave-guiding system with coupled unidirectional surface modes.

\end{document}